\documentclass[10pt]{article}
\usepackage{graphicx}
\usepackage{amsmath}
\usepackage{amssymb}
\usepackage{caption2}
\begin{document}
\bibliographystyle{prsty}
\begin{center}
{\large {\bf \sc{ Analysis of the vector meson transitions  among the heavy quarkonium  states  }}} \\[2mm]
Zhi-Gang Wang  \footnote{E-mail:wangzgyiti@yahoo.com.cn. } \\
  Department of Physics, North China Electric Power University, Baoding 071003, P. R.
  China
\end{center}

\begin{abstract}
In this article, we study the vector meson transitions among the
charmonium and bottomonium states with the heavy quark effective
theory in an systematic way, and make predictions for the ratios
among the vector meson decay widths of a special multiplet to
another multiplet. The predictions can be confronted with the
experimental data in the future.
\end{abstract}



\section{Introduction}

In 2003, the CLEO collaboration observed a  significant signal for
the transition $\Upsilon(3{\rm S}) \to \gamma \omega \Upsilon(1{\rm
S})$, which is consistent with the radiative decays $\Upsilon(3{\rm
S}) \to \gamma \chi_{b 1,2}(2{\rm P})$  followed by the hadronic
decays  $\chi_{b 1,2}(2{\rm P}) \to \omega \Upsilon(1{\rm S})$.
  The branching ratios are  ${\rm Br}\left(\chi_{b1}(2{\rm P}) \to \omega \Upsilon(1{\rm S})\right) =
  \left(1.63^{+0.35}_{ -0.31 }{}^{+0.16}_{-0.15}\right)\%$
   and ${\rm Br}\left(\chi_{b2}(2{\rm P}) \to \omega \Upsilon(1{\rm S})\right)
    = \left(1.10^{+0.32}_{-0.28}{}^{+0.11}_{-0.10}\right)\%$,
   respectively \cite{CLEO-Omega}.

   In 2004,  the Belle  collaboration observed a strong near-threshold
enhancement in the $\omega J/\psi$ invariant mass distribution  in
the exclusive $B\to K\omega J/\psi$ decays, the enhancement has a
mass of $(3943\pm 11\pm 13)\, {\rm MeV}$ and a total width of $( 87
\pm 22\pm 26)\, {\rm MeV}$ \cite{BelleY3940}. Later,  the Babar
collaboration confirmed the $Y(3940)$ in the exclusive decays
$B^{0,+}\to J/\psi\omega K^{0,+}$, the measured mass and width are
$\left(3914.6 ^{+3.8}_{-3.4} \pm{2.0}\right)\, \rm {MeV}$  and
$\left(34^{+12}_{-8}\pm{5}\right)\, \rm {MeV}$, respectively
\cite{BabarY3940}.  In 2009, the Belle collaboration reported the observation of a significant enhancement  with the mass $\left(3915 \pm 3 \pm 2 \right)\,\rm{MeV}$ and total width $ \left(17 \pm 10\pm 3\right)\,\rm{MeV} $ respectively in the process $\gamma \gamma \to \omega J/\psi$ \cite{Belle2010Y3940}, these values are
consistent with that of the $Y(3940)$.   The updated values of the  mass $\left(3919.1^{+3.8}_{-3.5} \pm 2.0\right)\,\rm{MeV}$  and total width $\left(31^{+10}_{-8}\pm 5\right)\,\rm{MeV}$ from the  Babar collaboration are also consistent with the old ones \cite{PRD2010Babar}.

In 2009,  the CDF collaboration observed a narrow structure
$Y(4140)$ near the $J/\psi\phi$ threshold with a statistical
significance in excess of $3.8\,\sigma$ in the exclusive $B\to
J/\psi\phi K$ decays produced in $\bar{p} p $ collisions
 \cite{CDF2009Y4140}. The measured mass and width are
$\left(4143.0\pm2.9 \pm1.2\right)\,\rm{MeV}$ and
$\left(11.7^{+8.3}_{-5.0}\pm3.7\right)\,\rm{MeV}$, respectively
\cite{CDF2009Y4140}. The Belle collaboration measured the process
$\gamma \gamma \to \phi J/\psi$ for the $ J/\psi\phi$ invariant mass
distributions, and observed a narrow peak $X(4350)$ with a
significance of $3.2\,\sigma$, and no signal for the $Y(4140)\to
J/\psi\phi$ structure was observed \cite{BelleX4350}. Recently, the
CDF collaboration confirmed the $Y(4140)$ in the $B^\pm\rightarrow
J/\psi\,\phi K^\pm$ decays  with a statistical significance greater
than $5\,\sigma$, the measured mass and width are
$\left(4143.4^{+2.9}_{-3.0} \pm0.6 \right)\, \rm{MeV}$ and
$\left(15.3^{+10.4}_{-6.1}\pm2.5\right)\,\rm{MeV}$, respectively
\cite{CDF1101}.

We can take the $Y(4140)$ as an exotic hybrid charmonium
 \cite{Y4140-hybrid} and the $Y(3940)$ as the
 $\chi_{c1}(2{\rm P})$ state \cite{BarnesIJMP} tentatively.
 The  $\chi_{c1}(2{\rm P})$ state has the dominant decay mode $D\bar{D}^*$
  and  a predicted width of
$140\,\rm{MeV}$, which is consistent with that of the $Y(3940)$
within the theoretical and experimental uncertainties
\cite{BarnesIJMP}. On the other hand, the decay $Y(3940) \to
D\bar{D}^*$ has not been observed yet, which disfavors such
identification. There have been several other identifications for
the $Y(4140)$  (the molecular state
\cite{Y4140-molecule,Y3940-4140}, the tetraquark state
\cite{Y4140-tetraquark}, the re-scattering effect \cite{Y4140-cc},
etc) and  the $Y(3940)$ (the tetraquark state \cite{T3872-3940}, the
molecular state \cite{Y3940-4140,Y3940-1}, etc).

If the $Y(3940)$ is really the $\chi_{c1}(2{\rm P})$ state,  the $2
\rm P$ charmonium and bottomonium  states have
 similar Okubo-Zweig-Iizuka suppressed decays. Experimentally, there is
 another $\chi_{c1}(2{\rm P})$ candidate, the $X(3872)$ \cite{cc3872}, which was observed  in the
$J/\psi\pi^{+}\pi^{-}$ invariant mass distribution  by the Belle
collaboration \cite{Belle0309032}, and  confirmed by the D0, CDF and
Babar collaborations \cite{ConfirmX3872CDF-D0-Babar}. If we take the
$Y(3940)$ as the $\chi_{c1}(2{\rm P})$ state, the $X(3872)$ has to
be assigned  to the molecular state \cite{M-H-3872,M3872}, the
hybrid state \cite{M-H-3872}, (not) the tetraquark state
(\cite{Not-T3872-1}) \cite{T3872-3940,T3872}, the threshold cusp
\cite{Cusp}, (not) the $\chi_{c1}(2{\rm P})$ state with some
$D\bar{D}^*+\bar{D}D^*$ component (\cite{Not-Couplie-1})
\cite{Couple-1}, etc.

In the past years, a number of charmonium-like states besides
 the $Y(3940)$, $Y(4140)$,  $X(3872)$ have been discovered, and many possible assignments for those
states have been suggested, such as the conventional charmonium
states, the multiquark states (irrespective of the molecule type and
the diquark-antidiquark type), the hybrid states, the baryonium
states, the threshold effects, etc \cite{Review-XYZ}. In this
article, we focus on the traditional charmonium and bottomonium
scenario, and do not mean such assignments are correct and exclude
other possibilities.

 In Ref.\cite{VoloshinOmega}, Voloshin assumes that  the hadronic decays $\chi_{b
1,2}(2{\rm P}) \to \omega \Upsilon(1{\rm S})$ take place through the
chromo-electric gluon fields ${\vec E}^a$,
\begin{eqnarray}
\langle \omega ({\vec\epsilon}) |d_{abc} \int  f(q_1, q_2, q_3){\vec
E}^a(q_1) \cdot {\vec E}^b(q_2)  {\vec E}^c(q_3)  \delta^{4} (q_1+
q_2+ q_3 - p) dq_1  dq_2  dq_3 |0 \rangle = A_\omega  {\vec
\epsilon}^* \, ,
\end{eqnarray}
 where $p=(m_\omega, {\vec 0})$, the ${\vec \epsilon}$ stands for the
polarization vector of the $\omega$ meson,  the $d_{abc}$ are the
symmetric $SU(3)$ constants, and the  $f(q_1, q_2, q_3)$ is a
totally symmetric form-factor.

The bottomonium state can emit three gluons in the $1^{--}$
channels, then the three gluons hadronize to the vector meson
$\omega$ or $\phi$, and the bottomonium state translates to another
bottomonium state subsequently.  The chromo-electric and
chromo-magnetic gluon fields ${\vec E}^a$ and ${\vec B}^a$ have the
quantum numbers $J^{PC}=1^{--}$ and $1^{+-}$, respectively.  The
chromo-magnetic gluon fields ${\vec B}^a$ are related to the heavy
quark spin-flipped transitions, and suppressed by the factors
$1/m_Q^n$ with $n\geq1$. The dominant contributions of the three
gluons to the light vector mesons come from the three
chromo-electric fields ${\vec E}^a$ with the special
 configuration $d_{abc}   {\vec E}^a \cdot {\vec E}^b \,\,{\vec
 E}^c$. We can integrate out the intermediate gluons, and obtain the
effective Lagrangians, which should obey the heavy quark symmetry.

If the initial and final heavy quarkonium states  have large energy
gaps, the emissions of three  energic  gluons are greatly
facilitated in the phase space and the corresponding decay widths
may be large, although such processes are Okubo-Zweig-Iizuka
suppressed in the flavor space and the strong coupling constant $|g_s|$ has smaller
value due to the larger energy scale. On the other hand, if the $\omega$
and $\phi$ transitions are kinematically suppressed in the phase
space, the branching ratios may be very small.  The gluons are
electro-neutral, the couplings of the special configuration $d_{abc}
{\vec E}^a \cdot {\vec E}^b \,\,{\vec E}^c$ to the photons are
supposed to be  very small and can be neglected. The radiative
transitions among the heavy quarkonium states can take place through
 the emissions of photons from the heavy quarks directly. The
dynamics which govern the transitions to the light vector mesons (through
three gluons) and to the photons are  different.
The  hadronic decays $\chi_{b 1,2}(2{\rm P}) \to \omega \Upsilon(1{\rm S})$  also receive
 contributions from the virtual photons through the vector meson dominance mechanism. In the
effective Lagrangians at the hadronic level, we do not need to distinguish the virtual
photon and three-gluon contributions,  which means the coupling constants in the effective
Lagrangians contain  contributions from both the electromagnetic and strong interactions.

The Okubo-Zweig-Iizuka suppressed decays can  take place through
the final-state re-scattering mechanism, for example,
\begin{eqnarray}
\chi_{b2}(2{\rm P})+q\bar{q} &\to& B^{*}(B)+\bar{B}^{*}(\bar{B}) \to
\Upsilon(1 {\rm S}) +\omega \, , \nonumber \\
\chi_{b1}(2{\rm P})+q\bar{q} &\to& B^{*}(B)+\bar{B}(\bar{B}^{*}) \to
\Upsilon(1 {\rm S}) +\omega \, .
\end{eqnarray}
If the masses of the initial heavy quarkonium states  are above or
near the thresholds of the  open-charmed or open-bottom meson pairs,
the transitions to the intermediate open-heavy mesons are
facilitated, there are  additional contributions from the final-state
re-scattering mechanism. The contributions from the intermediate open-charmed or open-bottom meson loops are
not necessary large compared with the tree level contributions. In Ref.\cite{FKGuo},  Guo et al perform systematic studies about
 the effects of the intermediate charmed meson loops in the $\pi^0$ and $\eta$ transitions among the charmonia.
  On the other hand, the intermediate meson loops can result
in mass shifts, continuum components and mixing amplitudes for the
conventional heavy quarkonium  states \cite{BianT-2}, the
$B^{*}\bar{B}^{*}$, $B\bar{B}$, $B^{*}\bar{B}$ and $B\bar{B}^{*}$
components can also result in the final-state $\Upsilon(1 {\rm
S})\omega$ through the exchange of the intermediate $B$ (or $B^*$)
meson. In such processes, the fusion of the open-heavy meson pairs
can result in a light vector meson or a photon, the corresponding
contributions are related with each other through the approximated
vector-meson dominance.

We can carry out the integral of  the intermediate meson-loops firstly, then
 parameterize the net effects by  some momentum-dependent
couplings of the $\chi_{b}\Upsilon\omega$ or $\chi_{c}J/\psi\omega$.
Here we prefer phenomenological analysis, and do not intend to obtain effective field theory, and do not
 separate the energy scales of the revelent degrees of freedom explicitly  so as to integrate out some of them in an systematic way.
The momenta  of the final states in the center of mass coordinate
are about $194\,\rm{MeV}$ and $135\,\rm{MeV}$ in the decays
$\chi_{b2}(2{\rm P}) \to \Upsilon(1 {\rm S})\omega$ and $
\chi_{b1}(2{\rm P})\to \Upsilon(1 {\rm S}) \omega $, respectively,
the vector meson $\omega$ is not energic. If we smear the
momentum-dependence of the coupling constants, we can obtain an
simple Lagrangian, which describes the Okubo-Zweig-Iizuka suppressed
$\omega$ or $\phi$ transitions among the heavy quarkonium states,
and should obey the heavy quark symmetry.

 In Ref.\cite{Wang1101}, we focus on the traditional charmonium and
bottomonium scenario and study the radiative transitions among the
charmonium and bottomonium states with the heavy quark effective
theory systematically. The charmonium and bottomonium states are
heavy quarkonium states, the heavy quark symmetry can put powerful
constraints  in diagnosing their natures. In this article, we extend
our previous works to study the vector-meson transitions among the
heavy quarkonium states based on the heavy quark effective theory
\cite{PRT1997,RevWise-Neubert}.

There are a number of unknown parameters which determine the
magnitudes of the scattering amplitudes in the multipole expansion
in QCD and the final-state re-scatterings mechanism. It is very
difficult to distinguish the contributions of the three-gluon
emissions from that of the open-heavy mesons re-scatterings
quantitatively without enough precise experimental data to fitting.
In this article, we smear the underlying dynamical details,
introduce momentum-independent
 coupling constants, and make estimations based on the heavy quark
symmetry.

The article is arranged as follows:  we study the vector meson
transitions among the heavy quarkonium states with the heavy quark
effective theory in Sect.2; in Sect.3, we present the
 numerical results and discussions; and Sect.4 is reserved for our
conclusions.

\section{ The vector meson transitions among the heavy quarkonium states }

The  heavy quarkonium states can be classified according to the
 notation ${\rm n}^{2s+1}L_{j}$, where the ${\rm n}$ is the radial quantum number, the $L$ is
  the orbital angular momentum, the $s$ is  the spin, and the $j$ is the total angular
momentum. They have the parity and charge conjugation $P=(-1)^{L+1}$
and $C=(-1)^{L+s}$, respectively. In the non-relativistic potential
quark models, the wave-functions $\psi(r,\theta,\varphi)$ of the
heavy quarkonium states can be written as
$R_{nL}(r)Y_{Lm}(\theta,\varphi)$ in the spherical coordinates,
where the $R_{nL}(r)$ are the radial wave-functions and the
$Y_{Lm}(\phi,\varphi)$ are the spherical harmonic functions.
 The states have the same radial quantum number ${\rm
n}$ and orbital momentum $L$ can be expressed by  the superfields
$J(n)$, $J^\mu(n)$, $J^{\mu\nu}(n)$, etc \cite{GattoPLB93},
\begin{eqnarray}
J&=&\frac{1+{\rlap{v}/}}{2}\left\{\Upsilon_{\mu}\gamma^\mu-\eta_b\gamma_5\right\}
\frac{1-{\rlap{v}/}}{2} \, , \nonumber \\
J^\mu&=&\frac{1+{\rlap{v}/}}{2}\left\{\chi_{2}^{\mu\nu}\gamma_\nu+\frac{1}{\sqrt{2}}\epsilon^{\mu\alpha\beta\lambda}v_\alpha
\gamma_{\beta}\chi^{1}_{\lambda}+\frac{1}{\sqrt{3}}\left(\gamma^\mu-v^\mu\right)\chi_{0}+h^\mu_b\gamma_5\right\}
\frac{1-{\rlap{v}/}}{2} \, , \nonumber \\
J^{\mu\nu}&=&\frac{1+{\rlap{v}/}}{2}\left\{\Upsilon_3^{\mu\nu\alpha}\gamma_\alpha+\frac{1}{\sqrt{6}}
\left[\epsilon^{\mu\alpha\beta\lambda}v_\alpha
\gamma_{\beta}g^{\tau\nu}+\epsilon^{\nu\alpha\beta\lambda}v_\alpha
\gamma_{\beta}g^{\tau\mu}\right]\Upsilon^2_{\tau\lambda}+\right.\nonumber\\
&&\left.\left[\sqrt{\frac{3}{20}}\left[\left(\gamma^\mu-v^\mu\right)g^{\nu\alpha}+\left(\gamma^\nu-v^\nu\right)g^{\mu\alpha}\right]-\frac{1}{\sqrt{15}}\left(g^{\mu\nu}-v^{\mu}v^{\nu}\right)\gamma^\alpha\right]\Upsilon_\alpha+\eta^{\mu\nu}_{2}\gamma_5\right\}
\frac{1-{\rlap{v}/}}{2} \, , \nonumber\\
\end{eqnarray}
where the $v^{\mu}$ denotes the four velocity associated to the
superfields. The superfields $J$, $J^\mu$, $J^{\mu\nu}$ are
functions of the radial quantum numbers ${\rm n}$, the fields in a
definite superfield have the same ${\rm n}$, and form a multiplet.
 Here (and subsequential) we write down the bottomonium states
explicitly, and smear the radial numbers $n$ in the fields for
simplicity, the corresponding ones for the charmonium states are
obtained with an simple replacement.  We multiply the bottomonium
fields $\Upsilon^3_{\mu\nu\alpha}$, $\Upsilon^2_{\mu\nu}$,
$\Upsilon_{\mu}$, $\eta^2_{\mu\nu}$, $\chi^2_{\mu\nu}$, $\cdots$
with a factor $\sqrt{M_{\Upsilon_3}}$, $\sqrt{M_{\Upsilon_2}}$,
$\sqrt{M_{\Upsilon}}$, $\sqrt{M_{\eta_2}}$, $\sqrt{M_{\chi_2}}$,
$\cdots$, and they have dimension of mass $\frac{3}{2}$.

The superfields $J^{\mu_1\ldots\mu_L}$ are completely symmetric,
traceless and orthogonal to the velocity, furthermore, they  have
the following properties under the parity, charge conjunction, heavy
quark spin transformations,
\begin{eqnarray}
J^{\mu_1\ldots\mu_L}&\stackrel{P}{\longrightarrow}&\gamma^{0}J_{\mu_1\ldots\mu_L}\gamma^{0} \, ,  \nonumber \\
J^{\mu_1\ldots\mu_L}&\stackrel{C}{\longrightarrow}&(-1)^{L+1}C[J_{\mu_1\ldots\mu_L}]^{T}C \,,\nonumber \\
J^{\mu_1\ldots\mu_L}&\stackrel{S}{\longrightarrow}&SJ_{\mu_1\ldots\mu_L}S^{\prime\dagger}\,,\nonumber \\
v^{\mu}&\stackrel{P}{\longrightarrow}&v_{\mu}\, ,
\end{eqnarray}
where $S,S^{\prime}\in SU(2)$ heavy quark spin symmetry groups, and
$[S,{\rlap{v}/}]=[S^{\prime},{\rlap{v}/}]=0$.

 The vector meson transitions between the ${\rm m}$ and
${\rm n}$ heavy quarkonium states  can be described by the following
Lagrangians,
\begin{eqnarray}
{\cal{L}}_{SS}&=&\sum_{m,n}\delta(m,n)\mathrm{Tr}\left[\bar{J}(m)\sigma_{\mu\nu}J(n)\right]F^{\mu\nu} \, , \nonumber\\
{\cal{L}}_{SP}&=&\sum_{m,n}\delta(m,n)\mathrm{Tr}\left[\bar{J}({m})J_{\mu}(n)+\bar{J}_\mu(n)
J(m)\right]V^{\mu} \, , \nonumber\\
{\cal{L}}_{PD}&=&\sum_{m,n}\delta(m,n)\mathrm{Tr}\left[\bar{J}_{\mu\nu}({m})J^{\nu}(n)+\bar{J}^\nu(n)
J_{\mu\nu}(m)\right]V^{\mu} \, ,
\end{eqnarray}
where $\bar{J}_{\mu_1\ldots\mu_L}=\gamma^0
J_{\mu_1\ldots\mu_L}^{\dag} \gamma^0$, $V^\mu=F^{\mu\nu}v_\nu$,
$F_{\mu\nu}=\sqrt{2}\left(\partial_\mu \omega_\nu-\partial_\nu
\omega_\mu\right)+\left(\partial_\mu \phi_\nu-\partial_\nu
\phi_\mu\right)$, and the $\delta(m,n)$ is the coupling constant.
The present Lagrangians are analogous  to the Lagrangians which
describe  the radiative transitions between the ${\rm m}$ and ${\rm
n}$ heavy quarkonium  states \cite{PRT1997,Wang1101,Fazio2008}.  The
Lagrangians ${\cal{L}}_{SP}$ and ${\cal{L}}_{PD}$ preserve parity,
charge conjugation, gauge invariance and heavy quark spin symmetry,
while the Lagrangian ${\cal{L}}_{SS}$ violates the heavy quark
symmetry.   The effective Lagrangians ${\cal{L}}_{SP}$ and
${\cal{L}}_{PD}$ describing the electric dipole $E_1$-like
transitions can be realized in the leading order $\mathcal {O}(1)$,
while the heavy quark spin violation effective Lagrangian
${\cal{L}}_{SS}$ describing the magnetic dipole $M_1$-like
transitions can be realized in the next-to-leading order $\mathcal
{O}(1/m_Q)$. In the heavy quark limit, the contributions of the
order $\mathcal {O}(1/m_Q)$ are greatly suppressed, and we expect
that the flavor and spin violation corrections of the order
$\mathcal {O}(1/m_Q)$ to the effective Lagrangians ${\cal{L}}_{SP}$
and ${\cal{L}}_{PD}$ are  smaller than (or not as large as) the
leading order contributions.

From the heavy quark effective Lagrangians  ${\cal L}_{SS}$, ${\cal
L}_{SP}$ and ${\cal L}_{PD}$, we can obtain the vector meson  decay
 widths $\Gamma$,
\begin{eqnarray}
\Gamma&=&\frac{1}{2j+1}\sum\frac{k_{V}}{8\pi M^2 } |T|^2\, ,
\end{eqnarray}
where the $T$ denotes the scattering amplitude, the $k_{V}$ is the
momentum of the final states in the center of mass coordinate, the
$\sum$ denotes the sum of all the  polarization  vectors,  the $j$
is the total angular momentum of the initial state, and the $M$ is
the mass of the initial state. The summation of the  polarization
vectors $\epsilon_{\mu}(\lambda,p)$, $\epsilon_{\mu\nu}(\lambda,p)$,
$\epsilon_{\mu\nu\rho}(\lambda,p)$ of the states with the total
angular momentum $j=1,2,3$ respectively
  results in the following three formulae,
 \begin{eqnarray}
 \sum_\lambda \epsilon^*_{\mu}\epsilon_{\nu}&=&\widetilde{g}_{\mu\nu}=-g_{\mu\nu}+\frac{p_\mu p_\nu}{p^2}  \, , \nonumber\\
 \sum_\lambda \epsilon^*_{\mu\nu}\epsilon_{\alpha\beta}&=&\frac{\widetilde{g}_{\mu\alpha}\widetilde{g}_{\nu\beta}
 +\widetilde{g}_{\mu\beta}\widetilde{g}_{\nu\alpha}}{2}-\frac{\widetilde{g}_{\mu\nu}\widetilde{g}_{\alpha\beta}}{3}  \, , \nonumber\\
\sum_\lambda\epsilon^*_{\mu\nu\rho}\epsilon_{\alpha\beta\tau}&=&\frac{1}{6}\left(
\widetilde{g}_{\mu\alpha}\widetilde{g}_{\nu\beta}\widetilde{g}_{\rho\tau}
+\widetilde{g}_{\mu\alpha}\widetilde{g}_{\nu\tau}\widetilde{g}_{\rho\beta}
+\widetilde{g}_{\mu\beta}\widetilde{g}_{\nu\alpha}\widetilde{g}_{\rho\tau}
   +\widetilde{g}_{\mu\beta}\widetilde{g}_{\nu\tau}\widetilde{g}_{\rho\alpha}
   +\widetilde{g}_{\mu\tau}\widetilde{g}_{\nu\alpha}\widetilde{g}_{\rho\beta}
   +\widetilde{g}_{\mu\tau}\widetilde{g}_{\nu\beta}\widetilde{g}_{\rho\alpha}\right)\nonumber\\
   &&-\frac{1}{15}\left(\widetilde{g}_{\mu\alpha}\widetilde{g}_{\nu\rho}\widetilde{g}_{\beta\tau}
   +\widetilde{g}_{\mu\beta}\widetilde{g}_{\nu\rho}\widetilde{g}_{\alpha\tau}
   +\widetilde{g}_{\mu\tau}\widetilde{g}_{\nu\rho}\widetilde{g}_{\alpha\beta}
   +\widetilde{g}_{\nu\alpha}\widetilde{g}_{\mu\rho}\widetilde{g}_{\beta\tau}
   +\widetilde{g}_{\nu\beta}\widetilde{g}_{\mu\rho}\widetilde{g}_{\alpha\tau}
      +\widetilde{g}_{\nu\tau}\widetilde{g}_{\mu\rho}\widetilde{g}_{\alpha\beta}\right. \nonumber\\
   &&\left. +\widetilde{g}_{\rho\alpha}\widetilde{g}_{\mu\nu}\widetilde{g}_{\beta\tau}
         +\widetilde{g}_{\rho\beta}\widetilde{g}_{\mu\nu}\widetilde{g}_{\alpha\tau}
         +\widetilde{g}_{\rho\tau}\widetilde{g}_{\mu\nu}\widetilde{g}_{\alpha\beta}\right)
         \, ,
 \end{eqnarray}
and we use the FeynCalc to carry out the contractions of the Lorentz
indexes.

\section{Numerical Results}

The heavy quarkonium states listed in the Review of Particle Physics
are far from complete and do not fill the spectroscopy \cite{PDG}.
Recently, the Belle collaboration  observed   the bottomonium states
$h_b(1\rm{P})$ and $h_b(2\rm{P})$  in the scattering  $e^+e^- \to
h_b({\rm{nP}})\pi^+\pi^-$ with significances of $5.5\,\sigma$ and
$11.2\,\sigma$, respectively \cite{Belle1103}. The measured masses
are $M_{h_b(1{\rm
{P}})}=\left(9898.25\pm1.06^{+1.03}_{-1.07}\right)\,\rm{MeV}$ and
$M_{h_b(2{\rm{P}})}=\left(10259.76\pm0.64^{+1.43}_{-1.03}\right)\,\rm{MeV}$,
respectively. There have been several theoretical works on the
spectroscopy of the charmonium and bottomonium states, such as the
relativized potential model (Godfrey-Isgur model)
\cite{Godfrey1985,Barnes2005}, the Cornell potential model, the
logarithmic potential model,
  the power-law potential model, the QCD-motivated potential model
\cite{Eichten1978,EQ1994}, the  relativistic quark model based on a
quasipotential approach in QCD \cite{Ebert2003},  the Cornell
potential model combined with heavy quark mass expansion
\cite{Roberts1995}, the screened potential model
\cite{LiChao2009,LiChao0909}, the potential non-relativistic QCD
model \cite{pNRQCD}, the confining  potential model with the
Bethe-Salpeter equation \cite{BSE},  etc.

In Tables 1-2, we  list
 the experimental values of the charmonium and  bottomonium
states compared with some theoretical predictions
\cite{PDG,Belle1103,Godfrey1985,Barnes2005,LiChao2009,LiChao0909}.
For the newly-observed charmonium-like states, there are hot
controversies about their natures, and we focus on the traditional
charmonium scenario, although such identifications are not superior
to others. One can consult Refs.\cite{Review-XYZ,Wang1101} for
detailed discussions.

\begin{table}
\begin{center}
\begin{tabular}{|cc|c|c|c|c|}
\hline\hline
 \multicolumn{2}{|c|}{State} &Experimental  \cite{PDG}          & Theoretical  \cite{LiChao2009}    & Theoretical  \cite{Barnes2005} &   Theoretical  \cite{Barnes2005} \\
\hline
1S &  $J/\psi(1^3{\rm S}_1) $      &   3096.916                  & 3097               & 3090 & 3098 \\
   &  $\eta_c(1^1{\rm S}_0) $      &   2980.3                    & 2979               & 2982 & 2975 \\
\hline
2S &  $\psi(2^3{\rm S}_1)  $       &   3686.09                   & 3673               & 3672 & 3676 \\
   &  $\eta_c(2^1{\rm S}_0)$       &   3637                      & 3623               & 3630 & 3623 \\
\hline
3S &  $\psi(3^3{\rm S}_1)   $      &   4039 $[\psi(4040)]$       & 4022               & 4072 & 4100 \\
   &  $\eta_c(3^1{\rm S}_0) $      &{\bf?}\,3942 $[X(3940)]$     & 3991               & 4043 & 4064 \\
\hline
4S &  $\psi(4^3{\rm S}_1)   $      & {\bf?}\,4263 $[Y(4260)]$    & 4273               & 4406 & 4450 \\
   &  $\eta_c(4^1{\rm S}_0) $      &                             & 4250               & 4384 & 4425 \\
\hline
5S &  $\psi(5^3{\rm S}_1)   $      & {\bf?}\,4421 $[\psi(4415)]$ & 4463               &      &      \\
   &  $\eta_c(5^1{\rm S}_0) $      &                             & 4446               &      &      \\
\hline
6S &  $\psi(6^3{\rm S}_1)   $      &{\bf?}\,4664 $[Y(4660)]$     & 4608               &      &      \\
   &  $\eta_c(6^1{\rm S}_0) $      &                             & 4595               &      &      \\
\hline
1P &  $\chi_{c2}(1^3{\rm P}_2) $   &  3556.20                    & 3554               & 3556 & 3550 \\
   &  $\chi_{c1}(1^3{\rm P}_1 )$   &  3510.66                    & 3510               & 3505 & 3510 \\
   &  $\chi_{c0}(1^3{\rm P}_0) $   &  3414.75                    & 3433               & 3424 & 3445 \\
   &  $h_c(1^1{\rm P}_1)    $      &  3525.42                    & 3519               & 3516 & 3517 \\
\hline
2P &  $\chi_{c2}(2^3{\rm P}_2) $   &  3929 $[Z(3930)]$           & 3937               & 3972 & 3979 \\
   &  $\chi_{c1}(2^3{\rm P}_1) $   &{\bf?}\,3914.6 $[Y(3940)]$   & 3901               & 3925 & 3953 \\
   &  $\chi_{c0}(2^3{\rm P}_0) $   &                             & 3842               & 3852 & 3916 \\
   &  $h_c(2^1{\rm P}_1) $         &                             & 3908               & 3934 & 3956 \\
\hline
3P &  $\chi_{c2}(3^3{\rm P}_2) $   &                             & 4208               & 4317 & 4337 \\
   &  $\chi_{c1}(3^3{\rm P}_1) $   &                             & 4178               & 4271 & 4317 \\
   &  $\chi_{c0}(3^3{\rm P}_0) $   &{\bf?}\,4156 $ [X(4160)]$    & 4131               & 4202 & 4292 \\
   &  $h_c(3^1{\rm P}_1) $         &                             & 4184               & 4279 & 4318 \\
\hline
1D &  $\psi_3(1^3{\rm D}_3) $      &                             & 3799               & 3806 & 3849 \\
   &  $\psi_2(1^3{\rm D}_2) $      &                             & 3798               & 3800 & 3838 \\
   &  $\psi(1^3{\rm D}_1) $        &   3772.92 $[\psi(3770)]$    & 3787               & 3785 & 3819 \\
   &$\eta_{c2}(1^1{\rm D}_2)$      &                             & 3796               & 3799 & 3837 \\
\hline
2D &  $\psi_3(2^3{\rm D}_3) $      &                             & 4103               & 4167 & 4217 \\
   &  $\psi_2(2^3{\rm D}_2) $      &                             & 4100               & 4158 & 4208 \\
   &  $\psi(2^3{\rm D}_1) $        &  4153 $[\psi(4160)]$        & 4089               & 4142 & 4194 \\
   &$\eta_{c2}(2^1{\rm D}_2)$      &                             & 4099               & 4158 & 4208 \\
\hline
3D &  $\psi_3(3^3{\rm D}_3) $      &                             & 4331               &      &      \\
   &  $\psi_2(3^3{\rm D}_2) $      &                             & 4327               &      &      \\
   &  $\psi(3^3{\rm D}_1) $        &{\bf?}\,4361 $[Y(4360)]$     & 4317               &      &      \\
   &$\eta_{c2}(3^1{\rm D}_2)$      &                             & 4326               &      &      \\
\hline\hline
\end{tabular}
\end{center}
 \caption{Experimental and theoretical mass spectrum of the
charmonium states, where the unit is MeV.}
 \end{table}

\begin{table}
\begin{center}
\begin{tabular}{|cc|c|c|c|}\hline
 \multicolumn{2}{|c|}{State}      & Experimental \cite{PDG,Belle1103} & Theoretical  \cite{LiChao0909}  & Theoretical  \cite{Godfrey1985}\\ \hline
1S &  $\Upsilon(1^3{\rm S}_1)$    & 9460.30                           & 9460                    & 9460    \\
   &  $\eta_b(1^1{\rm S}_0)$      & 9390.9                            & 9389                    & 9400    \\
\hline
2S &  $\Upsilon(2^3{\rm S}_1)$    & 10023.26                          & 10016                   & 10000   \\
   &  $\eta_b(2^1{\rm S}_0)$      &                                   & 9987                    & 9980    \\
\hline
3S &  $\Upsilon(3^3{\rm S}_1)$    & 10355.2                           & 10351                   & 10350   \\
   &  $\eta_b(3^1{\rm S}_0)$      &                                   & 10330                   & 10340   \\
\hline
4S &  $\Upsilon(4^3{\rm S}_1)$    & 10579.4                           & 10611                   & 10630   \\
   &  $\eta_b(4^1{\rm S}_0)$      &                                   & 10595                   &        \\
\hline
5S &  $\Upsilon(5^3{\rm S}_1)$    & 10865                             & 10831                   & 10880   \\
   &  $\eta_b(5^1{\rm S}_0)$      &                                   & 10817                   &        \\
\hline
6S &  $\Upsilon(6^3{\rm S}_1)$    & 11019                             & 11023                   & 11100   \\
   &  $\eta_b(6^1{\rm S}_0)$      &                                   & 11011                   &        \\
\hline
7S &  $\Upsilon(7^3{\rm S}_1)$    &                                   & 11193                   &        \\
   &  $\eta_b(7^1{\rm S}_0)$      &                                   & 11183                   &        \\
\hline
1P &  $\chi_{b2}(1^3{\rm P}_2)$   & 9912.21                           & 9918                    & 9900    \\
   &  $\chi_{b1}(1^3{\rm P}_1 )$  & 9892.78                           & 9897                    & 9880    \\
   &  $\chi_{b0}(1^3{\rm P}_0)$   & 9859.44                           & 9865                    & 9850    \\
   &  $h_b(1^1{\rm P}_1)$         & 9898.25                           & 9903                    & 9880    \\
\hline
2P &  $\chi_{b2}(2^3{\rm P}_2)$   & 10268.65                          & 10269                   & 10260   \\
   &  $\chi_{b1}(2^3{\rm P}_1)$   & 10255.46                          & 10251                   & 10250   \\
   &  $\chi_{b0}(2^3{\rm P}_0)$   & 10232.5                           & 10226                   & 10230   \\
   &  $h_b(2^1{\rm P}_1) $        & 10259.76                          & 10256                   & 10250   \\
\hline
3P &  $\chi_{b2}(3^3{\rm P}_2)$   &                                   & 10540                   &        \\
   &  $\chi_{b1}(3^3{\rm P}_1)$   &                                   & 10524                   &        \\
   &  $\chi_{b0}(3^3{\rm P}_0)$   &                                   & 10502                   &        \\
   &  $h_b(3^1{\rm P}_1)$         &                                   & 10529                   &        \\
\hline
4P &  $\chi_{b2}(4^3{\rm P}_2)$   &                                   & 10767                   &        \\
   &  $\chi_{b1}(4^3{\rm P}_1)$   &                                   & 10753                   &        \\
   &  $\chi_{b0}(4^3{\rm P}_0)$   &                                   & 10732                   &        \\
   &  $h_b(4^1{\rm P}_1)$         &                                   & 10757                   &        \\
\hline
5P &  $\chi_{b2}(5^3{\rm P}_2)$   &                                   & 10965                   &        \\
   &  $\chi_{b1}(5^3{\rm P}_1)$   &                                   & 10951                   &        \\
   &  $\chi_{b0}(5^3{\rm P}_0)$   &                                   & 10933                   &        \\
   &  $h_b(5^1{\rm P}_1)$         &                                   & 10955                   &        \\
\hline
1D &  $\Upsilon_3(1^3{\rm D}_3)$  &                                   & 10156                   & 10160   \\
   &  $\Upsilon_2(1^3{\rm D}_2)$  & 10161                             & 10151                   & 10150   \\
   &  $\Upsilon(1^3{\rm D}_1) $   &                                   & 10145                   & 10140   \\
   &$\eta_{b2}(1^1{\rm D}_2)$     &                                   & 10152                   & 10150   \\
\hline
2D &  $\Upsilon_3(2^3{\rm D}_3)$  &                                   & 10442                   & 10450   \\
   &  $\Upsilon_2(2^3{\rm D}_2)$  &                                   & 10438                   & 10450   \\
   &  $\Upsilon(2^3{\rm D}_1)$    &                                   & 10432                   & 10440   \\
   &$\eta_{b2}(2^1{\rm D}_2)$     &                                   & 10439                   & 10450   \\
\hline
3D &  $\Upsilon_3(3^3{\rm D}_3)$  &                                   & 10680                   &        \\
   &  $\Upsilon_2(3^3{\rm D}_2)$  &                                   & 10676                   &        \\
   &  $\Upsilon(3^3{\rm D}_1) $   &                                   & 10670                   &        \\
   &$\eta_{b2}(3^1{\rm D}_2)$     &                                   & 10677                   &        \\
\hline
4D &  $\Upsilon_3(4^3{\rm D}_3) $ &                                   & 10886                   &        \\
   &  $\Upsilon_2(4^3{\rm D}_2)$  &                                   & 10882                   &        \\
   &  $\Upsilon(4^3{\rm D}_1)$    &                                   & 10877                   &        \\
   &$\eta_{b2}(4^1{\rm D}_2)$     &                                   & 10883                   &        \\
\hline
5D &  $\Upsilon_3(5^3{\rm D}_3)$  &                                   & 11069                   &        \\
   &  $\Upsilon_2(5^3{\rm D}_2)$  &                                   & 11065                   &        \\
   &  $\Upsilon(5^3{\rm D}_1)$    &                                   & 11060                   &        \\
   &$\eta_{b2}(5^1{\rm D}_2)$     &                                   & 11066                   &        \\
\hline \hline
\end{tabular}
\caption{Experimental and theoretical mass spectrum of the
bottomonium states, where the unit is MeV. }
\end{center}
\end{table}

We calculate the vector meson decay widths $\Gamma$ using the
FeynCalc to carry out the contractions of the Lorentz indexes in the
summation of the polarization vectors. In calculations, the masses
of the charmonium and bottomonium states are taken as the
experimental values from the Particle Data Group \cite{PDG}, see
Tables 1-2; for the unobserved charmonium and bottomonium states, we
take the values from the screened potential model
\cite{LiChao2009,LiChao0909}.

\begin{table}
\begin{center}
\begin{tabular}{|c|c|c|c|c| }\hline\hline
                          & $\Gamma(\psi \to\eta_c \omega)$ $[k_V]$ & $\Gamma(\eta_c\to \psi\omega)$ $[k_V]$ & $\frac{\Gamma(\eta_c\to \psi\omega)}{\Gamma(\psi \to\eta_c \omega)}$   \\ \hline

                   $3{\rm S}\to1{\rm S}$       & 0.500  [616]    & 0.567 [283]    &  1.132  \\ \hline

                   $4{\rm S}\to1{\rm S}$       & 0.843  [858]    & 1.979 [728]    &  2.347  \\ \hline
         $\widehat{4{\rm S}\to1{\rm S}}$       & 0.386  [655]    & 0.763 [461]    &  1.976  \\ \hline

                   $5{\rm S}\to1{\rm S}$       & 1.113  [1007]   & 2.970 [927]    &  2.667  \\ \hline
         $\widehat{5{\rm S}\to1{\rm S}}$       & 0.548  [844]    & 1.410 [743]    &  2.570  \\ \hline

                   $6{\rm S}\to1{\rm S}$       & 1.579   [1215]  & 3.808 [1064]   &  2.412 \\ \hline
         $\widehat{6{\rm S}\to1{\rm S}}$       & 0.810   [1088]  & 1.901 [911]    &  2.348  \\ \hline
                   $6{\rm S}\to2{\rm S}$       & 0.494   [589]   & 0.927 [415]    &  1.878 \\ \hline \hline
\end{tabular}
\end{center}
\caption{ The ratios of the vector meson transitions of the $S$-wave
 to the $S$-wave charmonium states, where the
wide-hat denotes the corresponding $\phi$ transitions.   The units
of the widths and the $k_V$ are $\delta^2(m,n)$ and  MeV,
respectively. }
\end{table}

\begin{table}
\begin{center}
\begin{tabular}{|c|c|c|c|c| }\hline\hline
    & $\Gamma(\Upsilon \to\eta_b \omega)$ $[k_V]$ & $\Gamma(\eta_b\to \Upsilon\omega)$ $[k_V]$ & $\frac{\Gamma(\eta_b\to \Upsilon\omega)}{\Gamma(\Upsilon \to\eta_b \omega)}$   \\ \hline

                   $3{\rm S}\to1{\rm S}$       & 0.499   [537]     & 0.890  [363]     &  1.785  \\ \hline

                   $4{\rm S}\to1{\rm S}$       & 1.036   [844]     & 2.694  [777]     &  2.599  \\ \hline
         $\widehat{4{\rm S}\to1{\rm S}}$       & 0.410   [576]     & 0.954   [471]    &  2.326  \\ \hline

                   $5{\rm S}\to1{\rm S}$       & 1.940   [1164]    & 4.623   [1038]   &  2.383  \\ \hline
         $\widehat{5{\rm S}\to1{\rm S}}$       & 0.924   [991]     & 2.114   [838]    &  2.287  \\ \hline
                   $5{\rm S}\to2{\rm S}$       & 0.317   [382]     & 0.281   [127]    &  0.886  \\ \hline

                   $6{\rm S}\to1{\rm S}$       & 2.547   [1321]    & 6.721   [1244]   &  2.639  \\ \hline
         $\widehat{6{\rm S}\to1{\rm S}}$       & 1.248   [1174]    & 3.255   [1085]   &  2.609  \\ \hline
                   $6{\rm S}\to2{\rm S}$       & 0.655   [641]     & 1.667   [575]    &  2.545  \\ \hline

                   $7{\rm S}\to1{\rm S}$       & 3.339   [1492]    & 8.931   [1415]   &  2.675 \\ \hline
         $\widehat{7{\rm S}\to1{\rm S}}$       & 1.663   [1365]    & 4.422   [1280]   &  2.660 \\ \hline
                   $7{\rm S}\to2{\rm S}$       & 1.098   [868]     & 2.921   [811]    &  2.660  \\ \hline
         $\widehat{7{\rm S}\to2{\rm S}}$       & 0.445   [609]     & 1.096   [524]    &  2.464  \\ \hline
                   $7{\rm S}\to3{\rm S}$       & 0.285   [349]     & 0.604   [260]    &  2.119  \\ \hline \hline
\end{tabular}
\end{center}
\caption{ The ratios of the vector meson transitions of the $S$-wave
 to the $S$-wave bottomonium states, where the
wide-hat denotes the corresponding $\phi$ transitions. The units of
the widths and the $k_V$ are $\delta^2(m,n)$ and  MeV, respectively.
}
\end{table}

\begin{table}
\begin{center}
\begin{tabular}{|c|c|c|c|c|c| }\hline\hline
   $\Gamma$ & $\psi \to \chi_{2}\omega$ $[k_V]$ & $\psi \to \chi_{1}\omega$ $[k_V]$ & $\psi \to \chi_{0}\omega$ $[k_V]$ & $\eta_c \to h_c\omega$ $[k_V]$  \\ \hline

                   $4{\rm S}\to1{\rm P}$         &                 &                 & 1.668   [293]      &   \\ \hline

                   $5{\rm S}\to1{\rm P}$         & 9.712 [330]     & 7.669   [415]   & 3.747   [558]      & 24.238 [432] \\ \hline

                   $6{\rm S}\to1{\rm P}$         & 26.341  [688]   & 17.595  [739]   & 7.056   [839]      & 41.966 [641] \\ \hline
         $\widehat{6{\rm S}\to1{\rm P}}$         & 8.766 [379]     & 6.705   [469]   & 3.122   [620]      & 11.381 [284]  \\ \hline
                   $6{\rm S}\to2{\rm P}$         &                 &                 & 1.291   [228]      &           \\ \hline    \hline
\end{tabular}
\end{center}
\caption{ The  widths of the vector meson transitions of the
$S$-wave to the $P$-wave charmonium states, where the wide-hat
denotes the corresponding $\phi$ transitions. The units of the
widths and the $k_V$ are $10^{-2}\delta^2(m,n)$ and  MeV,
respectively. }
\end{table}

\begin{table}
\begin{center}
\begin{tabular}{|c|c|c|c|c|c| }\hline\hline
 $\Gamma$   & $\Upsilon \to \chi_{2}\omega$ $[k_V]$& $\Upsilon \to \chi_{1}\omega$ $[k_V]$& $\Upsilon \to \chi_{0}\omega$ $[k_V]$ & $\eta_b \to h_b\omega$ $[k_V]$ \\ \hline

                   $5{\rm S}\to1{\rm P}$         & 19.940  [519]    & 13.014  [551]   & 4.947 [602]   & 30.350   [460] \\ \hline

                   $6{\rm S}\to1{\rm P}$         & 34.853  [743]    & 22.122  [768]   & 8.074  [810]   & 63.730  [751]  \\ \hline
         $\widehat{6{\rm S}\to1{\rm P}}$         & 11.237  [409]    & 7.631   [454]   & 3.032  [523]   & 21.044  [423] \\ \hline
                   $6{\rm S}\to2{\rm P}$         &                  &                 & 0.455  [75]    &           \\ \hline

                   $7{\rm S}\to1{\rm P}$         & 55.030  [955]    & 34.489  [977]   & 12.319 [1015]  & 99.640  [960]  \\ \hline
         $\widehat{7{\rm S}\to1{\rm P}}$         & 23.994  [730]    & 15.238  [759]   & 5.563   [808]  & 43.626  [736]   \\ \hline
                   $7{\rm S}\to2{\rm P}$         & 17.464  [471]    & 11.186  [494]   & 4.141   [533]  & 31.230  [469]\\ \hline  \hline
\end{tabular}
\end{center}
\caption{ The  widths of the vector meson transitions of the
$S$-wave  to the $P$-wave bottomonium states, where the wide-hat
denotes the corresponding $\phi$ transitions. The units of the
widths and the $k_V$ are $10^{-2}\delta^2(m,n)$ and  MeV,
respectively.  }
\end{table}

\begin{table}
\begin{center}
\begin{tabular}{|c|c|c|c|c|c| }\hline\hline
  $\Gamma$  & $\chi_{2} \to \psi \omega$ $[k_V]$ & $ \chi_{1}\to \psi\omega$ $[k_V]$ & $\chi_{0} \to \psi\omega$ $[k_V]$ & $h_c \to \eta_c\omega$ $[k_V]$ \\ \hline

              $2{\rm P}\to1{\rm S}$      & 4.132 [251]      & 3.413 [211]      &                & 7.827   [436]     \\ \hline

              $3{\rm P}\to1{\rm S}$      & 14.918  [681]    & 13.788 [646]     & 12.863  [619]  & 17.935  [778]   \\ \hline
    $\widehat{3{\rm P}\to1{\rm S}}$      & 5.076  [380]     & 4.051  [310]     & 3.183   [248]  & 7.596  [542]   \\ \hline \hline
\end{tabular}
\end{center}
\caption{ The  widths of the vector meson transitions of the
$P$-wave  to the $S$-wave charmonium states, where the wide-hat
denotes the corresponding $\phi$ transitions. The units of the
widths and the $k_V$ are $10^{-2}\delta^2(m,n)$ and  MeV,
respectively. }
\end{table}

\begin{table}
\begin{center}
\begin{tabular}{|c|c|c|c|c|c| }\hline\hline
  $\Gamma$  & $\chi_{2} \to \Upsilon \omega$ $[k_V]$ & $ \chi_{1}\to \Upsilon\omega$ $[k_V]$ & $\chi_{0} \to \Upsilon\omega$ $[k_V]$ & $h_b \to \eta_b\omega$ $[k_V]$ \\ \hline

              $2{\rm P}\to1{\rm S}$      & 3.630   [194]   & 2.475   [135]   &                & 7.358  [361]   \\ \hline

              $3{\rm P}\to1{\rm S}$      & 19.053  [705]   & 18.105  [683]   & 16.804  [653]  & 22.599  [781]   \\ \hline
    $\widehat{3{\rm P}\to1{\rm S}}$      & 5.369   [337]   & 4.509   [288]   & 3.109   [203]  & 8.085   [478]   \\ \hline

              $4{\rm P}\to1{\rm S}$      & 34.608  [982]   & 33.565  [966]   & 31.897  [942]  & 39.146  [1048]   \\ \hline
    $\widehat{4{\rm P}\to1{\rm S}}$      & 15.382  [767]   & 14.770  [746]   & 13.843  [714]  & 17.981  [851]   \\ \hline

              $5{\rm P}\to1{\rm S}$      & 51.706  [1196]  & 50.475  [1182]  & 48.572  [1163] & 57.098  [1257]    \\ \hline
    $\widehat{5{\rm P}\to1{\rm S}}$      & 24.774  [1030]  & 24.084  [1012]  & 23.129  [990]  & 27.711  [1100]     \\ \hline
              $5{\rm P}\to2{\rm S}$      & 11.376  [501]   & 10.621  [477]   & 9.650   [444]  & 12.787  [544]  \\ \hline   \hline
\end{tabular}
\end{center}
\caption{ The  widths of the vector meson transitions of the
$P$-wave to the $S$-wave bottomonium states, where the wide-hat
denotes the corresponding $\phi$ transitions. The units of the
widths and the $k_V$ are $10^{-2}\delta^2(m,n)$ and  MeV,
respectively.  }
\end{table}

\begin{table}
\begin{center}
\begin{tabular}{|c|c|c|c|c|c|c|c|c| }\hline\hline
   $\Gamma$ & $\chi_{2} \to\Upsilon_3 \omega$& $ \chi_{2}\to \Upsilon_2\omega$ & $\chi_{2} \to \Upsilon\omega$ & $\chi_{1} \to \Upsilon_{2}\omega$ & $\chi_{1} \to \Upsilon\omega$ & $\chi_{0} \to \Upsilon\omega$ & $h_b \to \eta_2\omega$ \\
                             & $[k_V]$       & $[k_V]$      & $[k_V]$      & $[k_V]$    & $[k_V]$     & $[k_V]$     & $[k_V]$ \\ \hline
    $5{\rm P}\to1{\rm D}$    & 5.196         & 0.826        & 0.075        & 2.353      & 1.448       & 2.685       &  5.378    \\
                             & [197]         & [177]        & [235]        & [104]      & [185]       & [88]        &  [173]   \\ \hline
    \hline
\end{tabular}
\end{center}
\caption{ The  widths of the vector meson transitions of the
$P$-wave  to the $D$-wave bottomonium states. The units of the
widths and the $k_V$ are $10^{-2}\delta^2(m,n)$ and  MeV,
respectively.  }
\end{table}

\begin{table}
\begin{center}
\begin{tabular}{|c|c|c|c|c|c|c|c|c| }\hline\hline
 $\Gamma$   & $ \psi_3  \to \chi_{2}\omega$& $\psi_2\to  \chi_{2}\omega$ & $\psi_{2} \to  \chi_{1}\omega$ & $\psi \to \chi_{2}\omega$ & $\psi \to \chi_{1}\omega$ & $ \psi\to \chi_{0}\omega$ & $\eta_2 \to h_c\omega$ \\
                                        & $[k_V]$     & $[k_V]$     & $[k_V]$    & $[k_V]$       & $[k_V]$    & $[k_V]$   & $[k_V]$ \\ \hline

              $3{\rm D}\to1{\rm P}$    &              &             & 2.603      &  0.077        & 2.147      & 4.970     &  2.480    \\
                                       &              &             & [209]      &  [169]        & [299]      & [472]     &  [152]    \\ \hline \hline
\end{tabular}
\end{center}
\caption{ The  widths of the vector meson transitions of the
$D$-wave  to the $P$-wave charmonium states. The units of the widths
and the $k_V$ are $10^{-2}\delta^2(m,n)$ and  MeV, respectively. }
\end{table}

\begin{table}
\begin{center}
\begin{tabular}{|c|c|c|c|c|c|c|c|c| }\hline\hline
   $\Gamma$ & $ \Upsilon_3  \to \chi_{2}\omega$& $ \Upsilon_2\to  \chi_{2}\omega$ & $\Upsilon_{2} \to  \chi_{1}\omega$ & $\Upsilon \to \chi_{2}\omega$ & $\Upsilon \to \chi_{1}\omega$ & $ \Upsilon\to \chi_{0}\omega$ & $\eta_2 \to h_b\omega$ \\
                                       & $[k_V]$      & $[k_V]$     & $[k_V]$    & $[k_V]$      & $[k_V]$     & $[k_V]$    & $[k_V]$ \\ \hline

              $3{\rm D}\to1{\rm P}$    &              &             & 0.390      &              &             & 2.120      &       \\
                                       &              &             & [29]       &              &             & [203]      &       \\ \hline

              $4{\rm D}\to1{\rm P}$    & 13.116       & 3.225       & 10.464     & 0.350        & 5.699       & 8.622      &  13.705     \\
                                       & [553]        & [547]       & [577]      & [539]        & [569]       & [619]      &  [570]     \\ \hline

              $5{\rm D}\to1{\rm P}$    & 24.166       & 5.980       & 18.855     & 0.652        & 10.341      & 15.001     &  24.841      \\
                                       & [807]        & [802]       & [826]      & [795]        & [820]       & [860]      &  [820]     \\ \hline
    $\widehat{5{\rm D}\to1{\rm P}}$    & 8.999        & 2.206       & 7.244      & 0.239        & 3.935       & 6.036      &  9.466    \\
                                       & [518]        & [510]       & [547]      & [499]        & [538]       & [599]      &  [539]   \\ \hline
              $5{\rm D}\to2{\rm P}$    & 2.999        & 0.655       & 2.817      & 0.057        & 1.400       & 2.779      &  3.498    \\
                                       & [161]        & [142]       & [199]      & [113]        & [179]       & [259]      &  [186]    \\ \hline  \hline
\end{tabular}
\end{center}
\caption{ The  widths of the vector meson transitions of the
$D$-wave  to the $P$-wave bottomonium states, where the wide-hat
denotes the corresponding $\phi$ transitions. The units of the
widths and the $k_V$ are $10^{-2}\delta^2(m,n)$ and  MeV,
respectively. }
\end{table}

The numerical values of the vector meson transition widths  are
presented in Tables 3-11, where we retain the unknown coupling
constants $\delta(m,n)$
 among the multiplets of the radial quantum numbers $m$ and $n$.
 In general, we expect to fit
  the parameters $\delta(m,n)$  to the precise experimental data, however, in the
 present time the experimental data are rare. In Tables
 3-4,12-18, we present the ratios of the vector meson decay widths among the
 charmonium (and bottomonium) states.

 In Ref.\cite{VoloshinOmega}, Voloshin observes that the ratio of the vector meson decay widths
 $\chi_{b1,2}(2{\rm P})\to\Upsilon(1{\rm{S}})\omega$ can be approximated by the ratio of the $S$-wave
 phase factor,
 \begin{eqnarray}
  \frac{\Gamma(\chi_{b2}(2{\rm P})\to\Upsilon(1{\rm{S}})\omega)}
  {\Gamma(\chi_{b1}(2{\rm P})\to\Upsilon(1{\rm{S}})\omega)}&=&1.4\,(1.467)\, ,
   \end{eqnarray}
where the value in the bracket is the theoretical prediction  from
the heavy quark effective theory. The agreement  between the
approximated experimental data and the theoretical calculation based
on the heavy quark effective theory is rather good, and the heavy
quark effective theory works rather well. The ratios presented in
Tables 3-4,12-18 can be confronted with the experimental data in the
future at the BESIII, KEK-B, RHIC, $\rm{\bar{P}ANDA}$ and LHCb,  and
put powerful constraints  in identifying the $X$, $Y$, $Z$
charmonium-like (or bottomonium-like) mesons.

\begin{table}
\begin{center}
\begin{tabular}{|c|c|c|c|c|c| }\hline\hline
   $\widetilde{\Gamma}$ & $\psi \to \chi_{2}\omega$& $\psi \to \chi_{1}\omega$ & $\psi \to \chi_{0}\omega$ & $\eta_c \to h_c\omega$  \\ \hline

                   $5{\rm S}\to1{\rm P}$         & 1             & 0.790           & 0.386         & 2.496  \\ \hline

                   $6{\rm S}\to1{\rm P}$         & 1             & 0.668           & 0.268         & 1.593  \\ \hline
         $\widehat{6{\rm S}\to1{\rm P}}$         & 0.333         & 0.255           & 0.119         & 0.432   \\ \hline    \hline
\end{tabular}
\end{center}
\caption{ The ratios of the vector meson transitions of the $S$-wave
 to the $P$-wave charmonium states, where the
wide-hat denotes the corresponding $\phi$ transitions,
$\widetilde{\Gamma}=\frac{\Gamma}{\Gamma(\psi \to \chi_{2}\omega)}$.
}
\end{table}

\begin{table}
\begin{center}
\begin{tabular}{|c|c|c|c|c|c| }\hline\hline
   $\widetilde{\Gamma}$ & $\Upsilon \to \chi_{2}\omega$& $\Upsilon \to \chi_{1}\omega$ & $\Upsilon \to \chi_{0}\omega$ & $\eta_b \to h_b\omega$  \\ \hline

                   $5{\rm S}\to1{\rm P}$         & 1             & 0.653           & 0.248         & 1.522  \\ \hline

                   $6{\rm S}\to1{\rm P}$         & 1             & 0.635           & 0.232         & 1.829  \\ \hline
         $\widehat{6{\rm S}\to1{\rm P}}$         & 0.322         & 0.219           & 0.087         & 0.604  \\ \hline

                   $7{\rm S}\to1{\rm P}$         & 1             & 0.627           & 0.224         & 1.811   \\ \hline
         $\widehat{7{\rm S}\to1{\rm P}}$         & 0.436         & 0.277           & 0.101         & 0.793  \\ \hline
                   $7{\rm S}\to2{\rm P}$         & 1             & 0.641           & 0.237         & 1.788  \\ \hline  \hline
\end{tabular}
\end{center}
\caption{ The ratios of the vector meson transitions of the $S$-wave
 to the $P$-wave bottomonium states, where the
wide-hat denotes the corresponding $\phi$ transitions,
$\widetilde{\Gamma}=\frac{\Gamma}{\Gamma(\Upsilon \to
\chi_{2}\omega)}$. }
\end{table}

\begin{table}
\begin{center}
\begin{tabular}{|c|c|c|c|c|c| }\hline\hline
   $\widetilde{\Gamma}$ & $\chi_{2} \to \psi \omega$& $ \chi_{1}\to \psi\omega$ & $\chi_{0} \to \psi\omega$ & $h_c \to \eta_c\omega$  \\ \hline

              $2{\rm P}\to1{\rm S}$      & 1              & 0.826          &                & 1.894      \\ \hline

              $3{\rm P}\to1{\rm S}$      & 1              & 0.924          & 0.862          & 1.202    \\ \hline
    $\widehat{3{\rm P}\to1{\rm S}}$      & 0.340          & 0.272          & 0.213          & 0.509   \\ \hline \hline
\end{tabular}
\end{center}
\caption{ The ratios of the vector meson transitions of the $P$-wave
 to the $S$-wave charmonium states, where the
wide-hat denotes the corresponding $\phi$ transitions,
$\widetilde{\Gamma}=\frac{\Gamma}{\Gamma(\chi_{2} \to \psi
\omega)}$.   }
\end{table}

\begin{table}
\begin{center}
\begin{tabular}{|c|c|c|c|c|c| }\hline\hline
   $\widetilde{\Gamma}$ & $\chi_{2} \to \Upsilon \omega$& $ \chi_{1}\to \Upsilon\omega$ & $\chi_{0} \to \Upsilon\omega$ & $h_b \to \eta_b\omega$  \\ \hline

              $2{\rm P}\to1{\rm S}$      & 1              & 0.682          &                & 2.027    \\ \hline

              $3{\rm P}\to1{\rm S}$      & 1              & 0.950          & 0.882          & 1.186    \\ \hline
    $\widehat{3{\rm P}\to1{\rm S}}$      & 0.282          & 0.237          & 0.163          & 0.424    \\ \hline

              $4{\rm P}\to1{\rm S}$      & 1              & 0.970          & 0.922          & 1.131     \\ \hline
    $\widehat{4{\rm P}\to1{\rm S}}$      & 0.444          & 0.427          & 0.400          & 0.520    \\ \hline

              $5{\rm P}\to1{\rm S}$      & 1              & 0.976          & 0.939          & 1.104     \\ \hline
    $\widehat{5{\rm P}\to1{\rm S}}$      & 0.479          & 0.466          & 0.447          & 0.536      \\ \hline
              $5{\rm P}\to2{\rm S}$      & 1              & 0.934          & 0.848          & 1.124    \\ \hline   \hline
\end{tabular}
\end{center}
\caption{ The ratios of the vector meson transitions of the $P$-wave
 to the $S$-wave bottomonium states, where the
wide-hat denotes the corresponding $\phi$ transitions,
$\widetilde{\Gamma}=\frac{\Gamma}{\Gamma(\chi_{2} \to \Upsilon
\omega)}$.    }
\end{table}

\begin{table}
\begin{center}
\begin{tabular}{|c|c|c|c|c|c|c|c|c| }\hline\hline
   $\widetilde{\Gamma}$ & $\chi_{2} \to\Upsilon_3 \omega$& $ \chi_{2}\to \Upsilon_2\omega$ & $\chi_{2} \to \Upsilon\omega$ & $\chi_{1} \to \Upsilon_{2}\omega$ & $\chi_{1} \to \Upsilon\omega$ & $\chi_{0} \to \Upsilon\omega$ & $h_b \to \eta_2\omega$ \\ \hline

    $5{\rm P}\to1{\rm D}$    & 1         &  0.159       & 0.014      & 0.453      &  0.279     & 0.517     &  1.035    \\ \hline  \hline
\end{tabular}
\end{center}
\caption{ The ratios of the vector meson transitions of the $P$-wave
 to the $D$-wave bottomonium states,
$\widetilde{\Gamma}=\frac{\Gamma}{\Gamma(\chi_{2} \to\Upsilon_3
\omega)}$.   }
\end{table}

\begin{table}
\begin{center}
\begin{tabular}{|c|c|c|c|c|c|c|c|c| }\hline\hline
   $\widetilde{\Gamma}$ & $ \psi_3  \to \chi_{2}\omega$& $\psi_2\to  \chi_{2}\omega$ & $\psi_{2} \to  \chi_{1}\omega$ & $\psi \to \chi_{2}\omega$ & $\psi \to \chi_{1}\omega$ & $ \psi\to \chi_{0}\omega$ & $\eta_2 \to h_c\omega$ \\ \hline

              $3{\rm D}\to1{\rm P}$    &             &          & 1        & 0.030     & 0.825      & 1.910     & 0.953   \\ \hline \hline
\end{tabular}
\end{center}
\caption{ The ratios of the vector meson transitions of the $D$-wave
 to the $P$-wave charmonium states,
$\widetilde{\Gamma}=\frac{\Gamma}{\Gamma(\psi_{2} \to
\chi_{1}\omega)}$. }
\end{table}

\begin{table}
\begin{center}
\begin{tabular}{|c|c|c|c|c|c|c|c|c| }\hline\hline
 $\widetilde{\Gamma}$   & $ \Upsilon_3  \to \chi_{2}\omega$& $ \Upsilon_2\to  \chi_{2}\omega$ & $\Upsilon_{2} \to  \chi_{1}\omega$ & $\Upsilon \to \chi_{2}\omega$ & $\Upsilon \to \chi_{1}\omega$ & $ \Upsilon\to \chi_{0}\omega$ & $\eta_2 \to h_b\omega$ \\ \hline

              $3{\rm D}\to1{\rm P}$    &              &             & 1          &              &             & 5.435      &       \\ \hline

              $4{\rm D}\to1{\rm P}$    & 1            & 0.246       & 0.798      & 0.027        & 0.434       & 0.657      &  1.045   \\ \hline

              $5{\rm D}\to1{\rm P}$    & 1            & 0.247       & 0.780      & 0.027        & 0.428       & 0.621      &  1.028    \\ \hline
    $\widehat{5{\rm D}\to1{\rm P}}$    & 0.372        & 0.091       & 0.300      & 0.010        & 0.163       & 0.250      &  0.392  \\ \hline
              $5{\rm D}\to2{\rm P}$    & 1            & 0.218       & 0.939      & 0.019        & 0.467       & 0.927      &  1.166    \\ \hline  \hline
\end{tabular}
\end{center}
\caption{ The ratios of the vector meson transitions of the $D$-wave
 to the $P$-wave bottomonium states, where the
wide-hat denotes the corresponding $\phi$ transitions,
$\widetilde{\Gamma}=\frac{\Gamma}{\Gamma(\Upsilon_3 \to \chi_2
\omega)}$, while in the first line
$\widetilde{\Gamma}=\frac{\Gamma}{\Gamma(\Upsilon_{2} \to
\chi_{1}\omega)}$. }
\end{table}

In this article, we do not distinguish between the contributions of
the three-gluon emissions and  the open-heavy mesons re-scatterings,
 smear the underlying dynamical details, and introduce
momentum-independent coupling constants, which can be fitted to the
precise experimental data in the future. There is a relative
$P$-wave between the two final-state mesons, the decay widths
$\Gamma \propto k_V^3$, the uncertainties originate from the masses
 can be estimated as
\begin{eqnarray}
\frac{\Delta\Gamma}{\Gamma} &\approx& \frac{\Delta
k_V^3}{k_V^3}=3\frac{\Delta k_V}{k_V}\,.
\end{eqnarray}
The masses of the light vector mesons listed in the Review of
Particle Physics are $M_\phi=(1019.455\pm 0.020)\,\rm{MeV}$ and
$M_\omega=(782.65\pm 0.12)\,\rm{MeV}$ \cite{PDG}, the uncertainties
originate from the $\Delta M_{\omega/\phi}$ are tiny and can be
safely neglected. In calculations, we observe that  a larger $k_V$
results in a smaller uncertainty if the uncertainties of the masses
of the heavy quarkonium states are fixed. For example, the $k_V$ in
the transitions $\chi_{b1}({\rm 2P})\to \Upsilon({\rm 1S}) \omega$,
$\chi_{b2}({\rm 2P})\to \Upsilon({\rm 1S}) \omega$, $\chi_{b1}({\rm
3P})\to \Upsilon({\rm 1S}) \phi$,  $\chi_{b2}({\rm 3P})\to
\Upsilon({\rm 1S}) \phi$,  $\chi_{b1}({\rm 3P})\to \Upsilon({\rm
1S}) \omega$  and  $\chi_{b2}({\rm 3P})\to \Upsilon({\rm 1S})
\omega$ are $135\,\rm{MeV}$,  $194\,\rm{MeV}$, $288\,\rm{MeV}$,
$337\,\rm{MeV}$,  $683\,\rm{MeV}$  and $705\,\rm{MeV}$,
respectively, the uncertainty ($0.26\,\rm{MeV}$) of the
$M_{\Upsilon({\rm 1S})}$ results in  uncertainties about $3.2\%$,
$1.5\%$, $0.9\%$, $0.66\%$, $0.16\%$ and $0.15\%$, respectively. We
can obtain a qualitative estimation (large or small) about the
uncertainties based on the values of the $k_V$ presented in Tables
3-11.  In the case of large momentum transitions, for example,
$k_V\geq 500\,\rm{MeV}$, the uncertainties originate from the
 uncertainties of the heavy quarkonium  masses are very
 small (a few percent) and can be neglected, unless the uncertainties of the masses are large enough to
 distort the $k_V$ significantly. On the other hand, we should take
into account the mass uncertainties in the case of soft $k_V$ (about
$200\,\rm{MeV}$), where variations of the masses maybe result in
considerable uncertainties. At the present time, the heavy
quarkonium states listed in the Review of Particle Physics are far
from complete and do not fill the spectroscopy, we have to take the
masses from the quark models, where the uncertainties are usually
neglected,  detailed error analysis is beyond the present work. In this article,
we have neglected the corrections from  the terms  $\mathcal{O}(1/m^2_Q)$ and $\mathcal{O}(1/m_Q)$ for the Lagrangians
$\mathcal{L}_{SS}$ (as the spin-flipped Lagrangian $\mathcal{L}_{SS}$ is of order $\mathcal{O}(1/m_Q)$) and $\mathcal{L}_{SP}$, $\mathcal{L}_{PD}$ respectively in the heavy quark effective theory,
which maybe result in uncertainties larger that of the masses.

\section{Conclusion}
In this article, we study the vector meson transitions  among the
charmonium and bottomonium states with the heavy quark effective
theory in an systematic way, and make predictions for ratios among
the $\omega$ and $\phi$ decay widths of a special multiplet to
another multiplet, where the unknown couple constants $\delta(m,n)$
are canceled out with each other. The predictions can be confronted
with the experimental data in the future at the BESIII, KEK-B, RHIC,
$\rm{\bar{P}ANDA}$ and LHCb,  and put powerful  constraints  in
identifying  the $X$, $Y$, $Z$ charmonium-like (or bottomonium-like)
mesons.

\section*{Acknowledgment}
This  work is supported by National Natural Science Foundation of
China, Grant Number 11075053,  and the Fundamental Research Funds for the Central Universities.

\end{document}